\documentclass[aps,prl,twocolumn,footnoteinbib]{revtex4-1}

\usepackage{epsfig}
\usepackage{amssymb}
\usepackage{bm}
\usepackage{hyperref}
\usepackage{color}
\usepackage{amsmath}

\newcommand{\be}{\begin{equation}}
\newcommand{\ee}{\end{equation}}
\newcommand{\bea}{\begin{eqnarray}}
\newcommand{\eea}{\end{eqnarray}}

\newcommand{\degree}{^\circ}

\begin{document}

\title{Is there evidence for sterile neutrinos in IceCube data?}
\author{V.~Barger$^{1}$, Y.~Gao$^{2}$, D.~Marfatia$^{3,1}$\\}
\vspace{0.5cm}
\affiliation{%
\bigskip
$^1$Department of Physics, University of Wisconsin, Madison, WI 53706,
USA \\
$^2$Department of Physics, University of Oregon, Eugene, OR 97403, USA\\
$^3$Department of Physics and Astronomy, University of Kansas, Lawrence, KS
66045, USA}

\begin{abstract}
Data from the LSND and MiniBooNE experiments, and revised expectations of the antineutrino flux from 
nuclear reactors suggest the existence of eV-mass sterile neutrinos. $3+2$ and $1+3+1$ scenarios
accommodate all relevant short-baseline neutrino data except for the low-energy MiniBooNE anomaly.
We analyze the angular distribution of upward going atmospheric neutrino events in the IceCube-40 dataset for evidence
of sterile neutrinos within these scenarios. Depending on how systematic uncertainties are handled, we find strong
evidence for, or weak evidence against sterile neutrinos. We show that future IceCube data will definitively settle the issue.

\end{abstract}

\maketitle

While oscillations of the three Standard Model neutrinos have been convincingly demonstrated, and
the framework that explains the data of many disparate experiments is elegant in its simplicity, a nagging doubt
persists. Are we seeing the effects of light sterile neutrinos in short-baseline neutrino (SBL) experiments? The question is
raised by 3.3$\sigma$ CL evidence from the Liquid Scintillator Neutrino Detector (LSND)~\cite{Aguilar:2001ty} experiment for $\bar{\nu}_\mu\rightarrow \bar{\nu}_e$ oscillations with $L/E \sim 1$~m/MeV. This result is partially
corroborated by the Mini-Booster Neutrino Experiment (MiniBooNE)~\cite{AguilarArevalo:2010wv}.
The mass-squared difference $\delta m^2 \sim 1$ eV$^2$ indicated is very different from those that explain solar and
atmospheric neutrino data. Since three neutrinos have only two independent mass-squared difference scales, a third
$\delta m^2$ scale implies the existence of additional light neutrinos, which must be sterile because the 
invisible width of the $Z$ boson requires that there be no more than three active neutrinos.

A new theoretical calculation of the $\bar{\nu}_e$ flux from nuclear reactors finds  a value
that is 3\% higher than previous estimates~\cite{Mueller:2011nm}. If borne out, the incident flux at 
SBL reactor neutrino experiments may be interpreted as having a deficit due to oscillations into
eV sterile neutrinos.

A recent analysis of SBL beam and reactor experiments, using the new reactor flux prediction, concludes
that a $3+1$ scenario with a single additional eV sterile neutrino does not describe the data satisfactorily, while a $3+2$ scenario with two eV sterile neutrinos, and a $1+3+1$ scenario with one sterile neutrino lighter than the 3 active neutrinos and the other heavier, provide a good fit; see Table~\ref{tab:maltoni} for the best-fit parameters~\cite{Kopp:2011qd}.
The phase $\delta\equiv arg(U_{\mu 4}U^*_{e 4}U^*_{\mu 5}U_{e 5})$, where $U$ is the neutrino mixing matrix, leads to CP violation.
Note that accommodating such eV sterile neutrinos in cosmology requires quite drastic modifications
of $\Lambda$CDM~\cite{Hamann:2010bk}.

\begin{table}
\begin{tabular}{c|cc|cccc|c}
\hline
&$\delta m^2_{41}$\ & \ $\delta m^2_{51}$\  & \ $|U_{e4}|$\  & \ $|U_{\mu 4}|$\  & \ $|U_{e5}|$\  & \ $|U_{\mu 5}|$\ &\ $\delta/\pi$\   \\
\hline
3+2&0.47&0.87 & 0.128& 0.165 &0.138 &0.148 &1.64 \\
1+3+1(a)& -0.47 &0.87& 0.129&0.154&0.142&0.163& 0.35\\
1+3+1(b)& 0.47& -0.87& 0.129&0.154&0.142&0.163& 1.65\\
\hline
\end{tabular}
\caption{Global best-fit parameters to data from short-baseline experiments~\cite{Kopp:2011qd}. 
The two $1+3+1$ cases correspond to either $m_4$ or $m_5$ being the lightest state. 
The active neutrinos have a normal hierarchy, and $\theta_{13}=0$. 
}
\label{tab:maltoni}
\end{table}

The IceCube experiment (IC)~\cite{Gaisser:2011iz}  observes 10$^5$
atmospheric muon neutrino events per year thereby offering a unique 
probe of neutrino oscillations at energies above 100~GeV.
For upward going neutrinos, any additional species that significantly mixes
with $\nu_\mu$ impacts the observed muon rate through $\nu_\mu$ disappearance. 
In fact, for $\delta m^2 \sim 1$~eV$^2$, resonant oscillations are possible for TeV atmospheric 
neutrinos~\cite{Ribeiro:2007ud}.
In its 40-string configuration, IC has observed 12877 atmospheric muon neutrino events in the energy range 
332~GeV--84~TeV~\cite{Abbasi:2011jx}.  

\begin{figure}
\includegraphics[width=1.06\columnwidth]{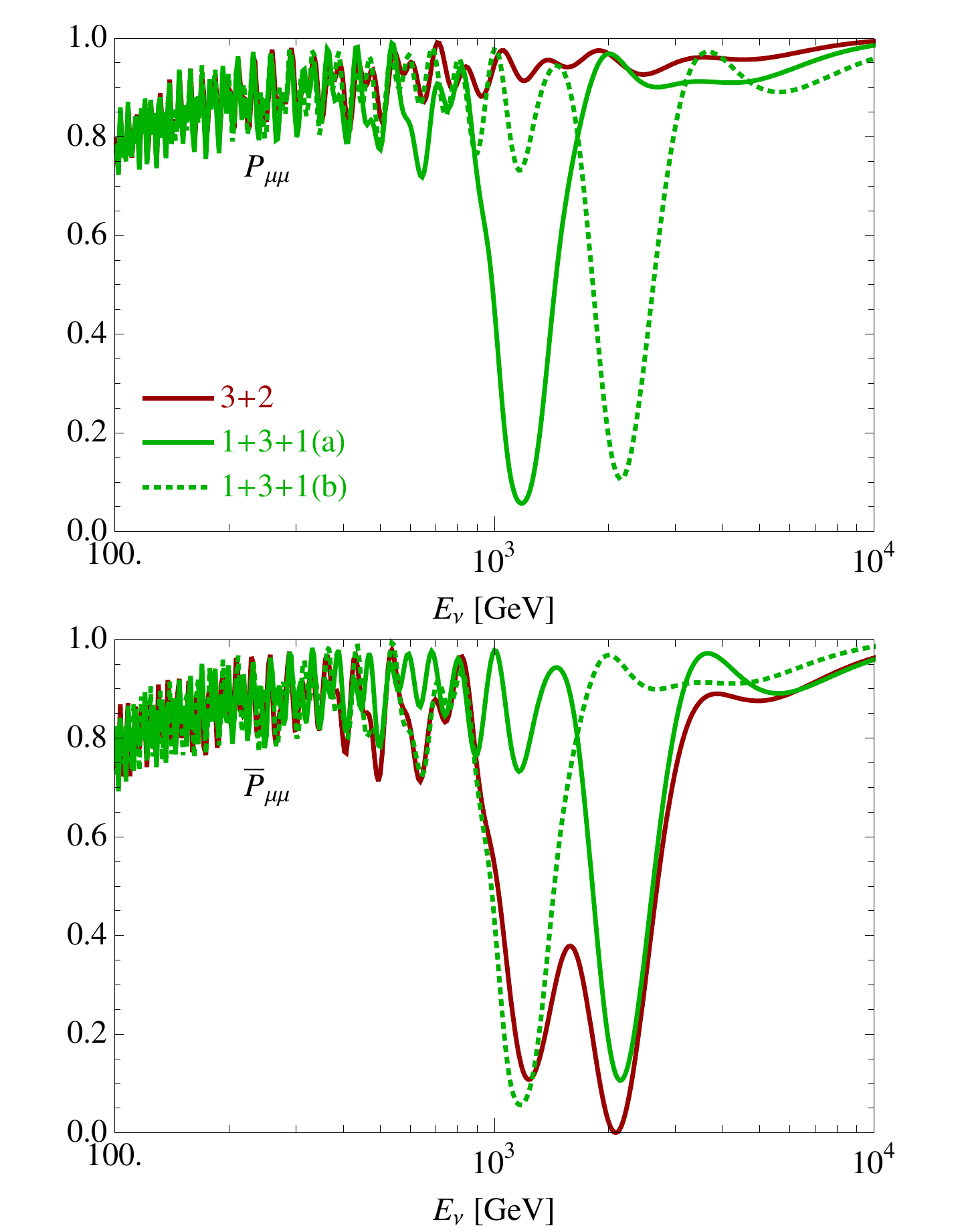}
\caption{The survival probability for $\nu_\mu$ (upper panel) and $\bar{\nu}_\mu$ (lower panel) for vertically going upward neutrinos \mbox{($\cos\theta_z=-1$)} in the 3 scenarios of Table~\ref{tab:maltoni}.  We use the density profile of the
Preliminary Reference Earth Model~\cite{Dziewonski:1981xy}. }
\label{fig:psurvival}
\end{figure}

In this Letter, we analyze the zenith-angle distribution of the IC data to see if it provides supporting evidence for sterile neutrino explanations
of the SBL data and the reactor antineutrino anomaly as encapsulated by the scenarios of Table~\ref{tab:maltoni}, and assess the future sensitivity of the IC detector.
The energy window of the IC dataset is optimal as can be seen from Fig.~\ref{fig:psurvival}.  We expect IC data to not be very sensitive to the $3+2$ case
since the resonance occurs in the $\bar{\nu}_\mu$ channel which is subdominant in the atmospheric flux. 
In the $1+3+1$ cases, the dominant $\nu_\mu$ flux is also suppressed so that these cases
can be discriminated more readily from the $3\nu$ case. 

The muon events at IC can be classified as (i) `contained' events, namely events
 with muon tracks that start within the instrumented volume,
 and (ii) `up-going' events in which the muon is produced outside the detector. See Eqs.~8 and~10 of 
 Ref.~\cite{Barger:2011em} for details and note that
 the incident $\nu_\mu$ flux includes a contribution from $\nu_e \to \nu_\mu$ oscillations.
 



For the points in Table~\ref{tab:maltoni}, the muon neutrino flux is suppressed by $\sim 10$\% above 100~GeV, with a corresponding reduction in the muon event rate.  Thus, the 25\% uncertainty in 
the atmospheric neutrino flux~\cite{Honda:2006qj} is a serious impediment to a rate analysis. Instead, a distortion in 
the zenith-angle distribution of the detected muons would
provide the strongest evidence for sterile neutrinos. 
As most sub-TeV neutrinos do not distort the angular distribution, useful information 
may be extracted from high-energy events. It so happens that the cut at 332~GeV suppresses 90\% of
 sub-TeV muon events while leaving a large sample of $10^4$ events. 

To make it to the detector, up-going muons can travel at most the stopping distance which increases with energy. 
As a result, the effective volume for neutrinos is larger for high energy neutrinos that 
more readily produce energetic muons.  
In turn, the up-going sample has a larger fraction of
high-energy events than the contained sample, making it particularly valuable in an analysis of the zenith-angle
distribution. The 40-string IC data consist of 70\% up-going and 30\% contained events~\cite{sean}.

\begin{figure}[t]
\includegraphics[width=1\columnwidth]{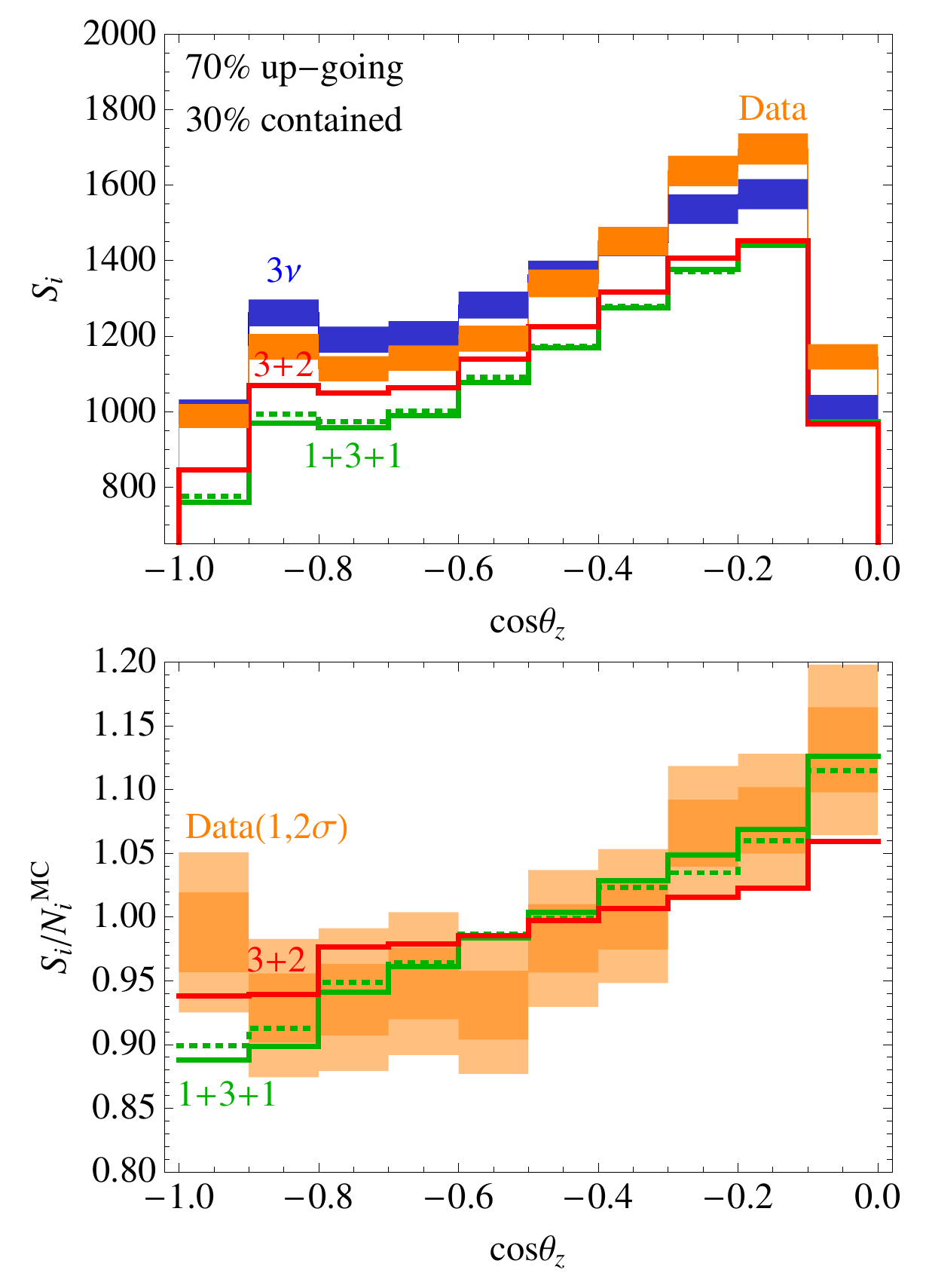}
\caption{Upper panel: Zenith-angle distributions for muon events at IC. 
The IC data (light, orange) and the expectation for $3\nu$ oscillations (dark, blue) are shown with $2\sigma$
statistical uncertainty bands. 
The expectations for the $3+2$ (dark, red), $1+3+1$(a)  (light, green solid), and $1+3+1$(b) (light, green dotted)
scenarios are shown. We set $a=1\,,b=0$ to display the rate suppression and the spectral shape in each case.
Lower panel: Relative distortion of the zenith-angle spectra in comparison to the expectation for 
$3\nu$ oscillations. Here, the
best-fit value of $a$ is used in each case, but $b=0$. The data reveal an obvious distortion with respect to the
$3\nu$ expectation.
}
\label{fig:40string}
\end{figure}

To account for experimental efficiencies, we determine the theoretical zenith-angle spectrum as 
in Ref.~\cite{Razzaque:2011ab}:
\be 
S_i^{th}=a[1+b((\cos\theta_z)_i+0.5)]{\frac{N^{MC}_i}{N^{3\nu}_{i}}N_i^{th}}\,,
\label{eq:scaling}
\ee
where $N_i^{th}$ is the number of muon events in angular bin $i$, and 
 $N^{MC}_i/N^{3\nu}_{i}$ is a bin-wise factor that
scales our theoretical $3\nu$ prediction to IceCube's Monte Carlo~\cite{Abbasi:2011jx}. 
$a$ is an overall normalization and $b$ allows a systematic tilt of the spectrum. Both parameters are
 allowed to float in a $\chi^2$ analysis.
Note that $S^{th}_i=N^{MC}_{i}$ in the $3\nu$ case with $a=1\,, b=0$.

The upper panel of Fig.~\ref{fig:40string} illustrates the event rate suppression and
zenith-angle distributions in the scenarios under consideration. 
The large difference of the sterile neutrino scenarios from the IC data and from the $3\nu$ expectation in the near-vertical ($-1.0\leq \cos\theta_z \leq -0.9$) bin promises much discriminating power.
The near-horizon ($-0.1\leq \cos\theta_z \leq 0$) bin
has potentially large systematic errors from the misidentification of coincident downward events as 
horizontal events, and we do not include it in our $\chi^2$ analysis.
For our analyses we define
\be 
\chi^2=\frac{(1-a)^2}{\sigma_a^2} + \sum_{i}^{}\frac{(S_i^{th}-S_i^{exp})^2}{S_i^{exp}}\,,
\label{eq:ang_chi2}
\ee
where $\sigma_a=0.25$ is the percent uncertainty in the atmospheric flux normalization~\cite{Honda:2006qj}  and $S^{exp}$ denotes either real or simulated data. 
In what follows, we always fit $a$, and either set $b=0$ or marginalize over $b$ without penalty. 
The lower panel of Fig.~\ref{fig:40string} shows the spectral distortion of the IC data and the sterile neutrino scenarios relative to $3\nu$ oscillations. In
the latter case, $a=1\,,b=0$, while the best-fit value of $a$ is used for the sterile cases with $b=0$.

\begin{figure}[t]
\includegraphics[width=0.9\columnwidth]{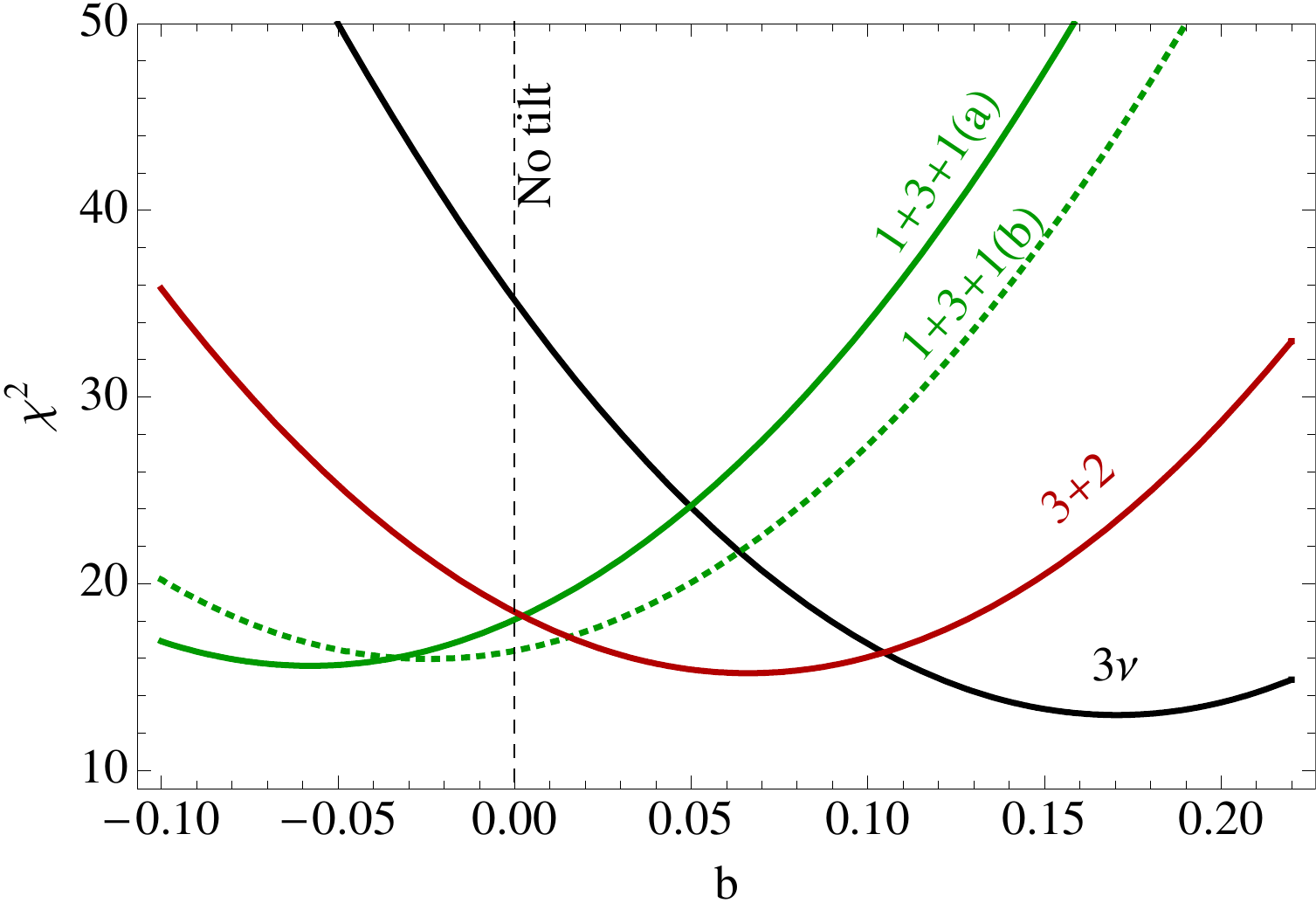}
\caption{$\chi^2$ as a function of the tilt parameter $b$ from a fit to the 9 zenith-angle bins of Fig.~\ref{fig:40string}
 with $\cos\theta_z<-0.1$.
}
\label{fig:bestfit}
\end{figure}

The results of fitting the 9
non-horizontal ($\cos\theta_z<-0.1$) bins are shown in Fig.~\ref{fig:bestfit} and Table~\ref{tab:chi2}.
The $3\nu$ scenario gives a better fit provided the data are plagued by a large systematic tilt. 
If $|b|$ is restricted to
be smaller than 0.11, the $3\nu$ fit yields a $\chi^2$ of 15.8, which is comparable to the values ($\sim$~15--16) of
 the sterile neutrino cases.
 
 A natural question is if there are correlated signals in cascade events at IC. 
  Interestingly, above 332~GeV, we find a comparable total number of $\nu_e$ plus $\nu_\tau$ 
  events/km$^3$/year in the $3\nu$ ($\sim 890$) and sterile neutrino ($\sim 850$) cases. The corresponding number of induced cascade events can be obtained by folding with the IC efficiency which is not available to us.

\begin{table}[t]
\begin{tabular}{c|c|cc}
\hline
		&no tilt ($b=0$)&\ \ \ \ \ \ \ \ \ \ \ tilted &\\
\hline
                   &  $\chi^2$        &  $\chi^2$ & $b$\\
 \hline                  
$3\nu$ 	&35.2	&13.0	& 0.17	\\
$3+2$		&18.5	&15.2	& 0.066 \\
$1+3+1$(a)&18.1	&15.6	& -0.058\\
$1+3+1$(b)&16.4	&16.0	&  -0.024\\
\hline
\end{tabular}
\caption{
Best-fit $\chi^2$ and $b$ from the curves in Fig.~\ref{fig:bestfit}.  Without a systematic tilt 
in the angular distribution, sterile neutrino scenarios are clearly favored over the $3\nu$ case. If a large 
systematic tilt is permissible, the sterile cases are mildly disfavored. For $|b|<0.11$, all scenarios give a comparable fit.}
\label{tab:chi2}
\end{table}



\begin{figure}[t]
\includegraphics[width=1.\columnwidth]{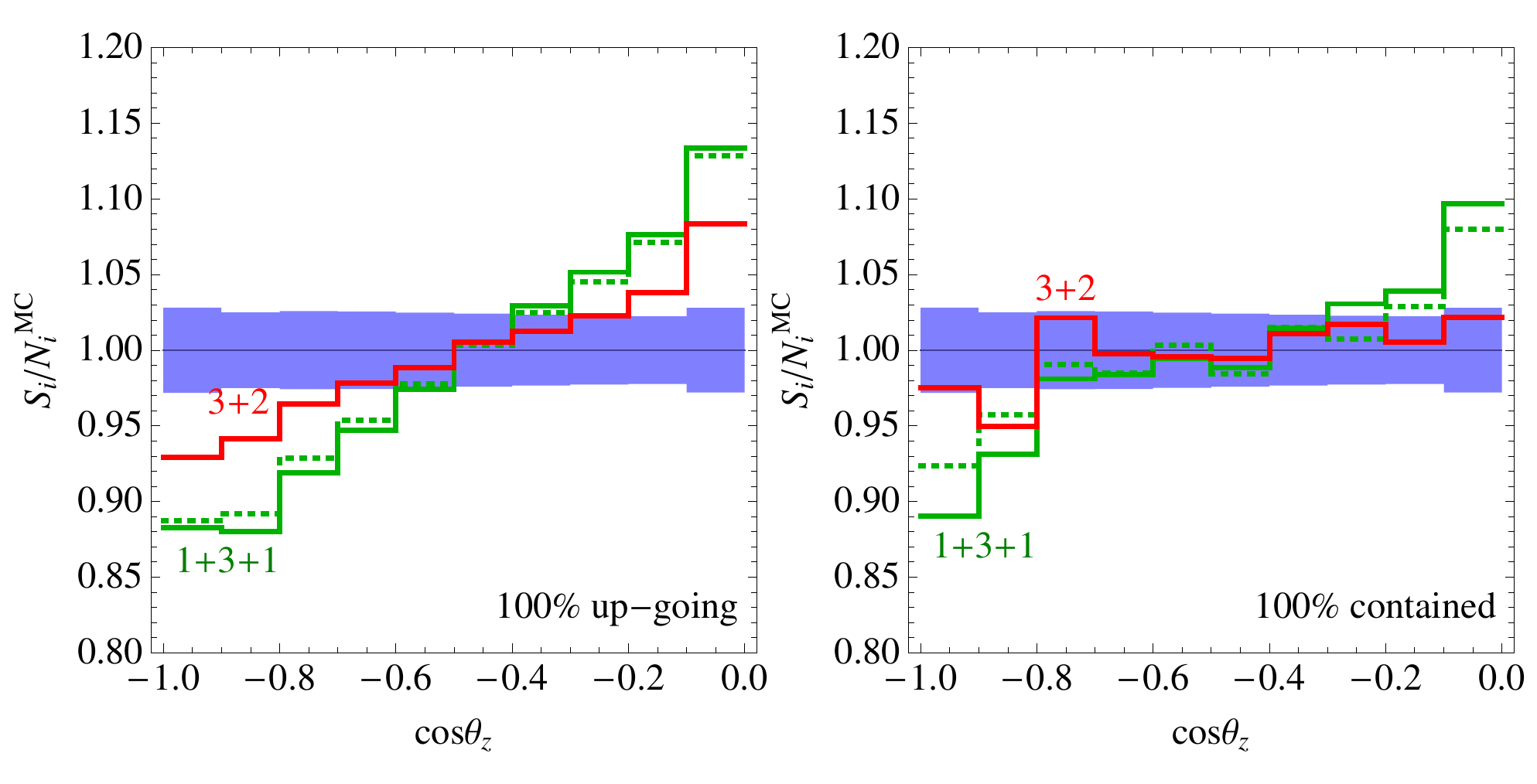}
\caption{
Similar to the lower panel of Fig.~\ref{fig:40string}, but 
with \mbox{$1.3\times 10^5$} events above 332 GeV equipartitioned into purely up-going events (left panel)
and purely contained events (right panel). The shaded band represents the $2\sigma$ statistical uncertainty on the $3\nu$ expectation. Sterile neutrino scenarios will be easily distinguishable from $3\nu$ oscillations from the up-going event sample.
}
\label{fig:future}
\end{figure}

That future IC data with systematics under control have
the ability to reveal a deviation due to sterile neutrinos is evident from
 Fig.~\ref{fig:future}. The monotonically rising event ratio as a function of $\cos\theta_z$ in
 the up-going event sample is a striking signature of sterile neutrino oscillations. 
 On the other hand, strong exclusions can also be obtained.
Assuming a sample of $6.5\times 10^4$ up-going events with no deviation
from the $3\nu$ result, and fitting to all 10 angular bins with only $a$ varied yields
$\chi^2=111$, 386 and 331 for the $3+2$, $1+3+1$(a) and (b) cases, respectively, compared to a $\chi^2 \sim 10$
for $3\nu$ oscillations. An interesting aspect of the contained event sample is that
in all cases the zenith-angle distribution is almost flat for $-0.8 < \cos\theta_z < -0.2$ so that the 
 features in the near-vertical and near-horizon bins are insensitive to a systematic tilt.
With  $6.5\times 10^4$ contained events, we find $\chi^2=28$, 165 and 87 
for the $3+2$ and  $1+3+1$(a) and (b) cases, respectively. With such sensitivity IC will easily
confirm or exclude sterile neutrino scenarios.

\vspace{0.1cm}
{\it Acknowledgements.}
We thank S. Grullon, F. Halzen, W.~Huelsnitz, J.~Koskinen, B. Louis, S. Pakvasa and T. Schwetz  for useful discussions and correspondence. VB thanks the KITP, UCSB for its hospitality. DM thanks the University of Hawaii for its hospitality during the completion of this work. This research was supported by 
DOE grants DE-FG02-96ER40969, DE-FG02-95ER40896 and DE-FG02-04ER41308,
and NSF grants PHY-0544278 and PHY05-51164.


\label{bib}

\end{document}